\documentclass[]{article}	

\usepackage[utf8]{inputenc}	
\usepackage[big,online]{dgruyter}	
\usepackage{lmodern} 
\usepackage{microtype}
\usepackage{biblatex}
\bibliography{bibliography.bib}
\usepackage{csquotes}
\usepackage{comment}
\usepackage{wrapfig}

\newcommand{\eg}{\textit{e.g.},\ }
  \newcommand{\ie}{\textit{i.e.},\ }

\theoremstyle{dgthm}

\theoremstyle{dgdef}

\begin{document}

  \journalname{}
  \journalyear{2024}

\title{Quantifying and combining uncertainty for improving the behavior of Digital Twin Systems}
\runningtitle{Quantifying Uncertainty in Digital Twin Systems}

\author[1]{Julien Deantoni}
\author[2]{Paula Muñoz}
\author[3]{Cláudio Gomes}
\author[4]{Clark Verbrugge}
\author[5]{Rakshit Mittal}
\author[6]{Robert Heinrich}
\author[5]{Stijn Bellis}
\author[2]{Antonio Vallecillo}
\runningauthor{J.~Deantoni et al.}
\affil[1]{\protect\raggedright 
Universite Cote d'Azur, France, e-mail: julien.deantoni@univ-cotedazur.fr}
\affil[2]{\protect\raggedright 
ITIS Software, Universidad de M\'alaga, Spain, e-mail: paulam@uma.es, av@uma.es}
\affil[3]{\protect\raggedright 
Aarhus University, Denmark, e-mail: claudio.gomes@ce.au.dk}
\affil[4]{\protect\raggedright 
McGill University, Canada, e-mail: clump@cs.mcgill.ca}
\affil[5]{\protect\raggedright 
University of Antwerp - Flanders Make, Belgium, e-mail: rakshit.mittal@uantwerpen.be, stijn.bellis@uantwerpen.be}
\affil[6]{\protect\raggedright 
Karlsruhe Institute of Technology, Germany, e-mail: robert.heinrich@kit.edu}

\abstract{ Uncertainty is an inherent property of any complex system, especially those that integrate physical parts or operate in real environments. In this paper, we focus on the Digital Twins of adaptive systems, which are particularly complex to design, verify, and optimize. One of the problems of having two systems (the physical one and its digital replica) is that their behavior may not always be consistent. In addition, both twins are normally subject to different types of uncertainties, which complicates their comparison. In this paper we propose the explicit representation and treatment of the uncertainty of both twins, and show how this enables a more accurate comparison of their behaviors. Furthermore, this allows us to reduce the overall system uncertainty and improve its behavior by properly averaging the individual uncertainties of the two twins. An exemplary incubator system is used to illustrate and validate our proposal.}

\keywords{Model-based Software Engineering, Control systems, Self-adaptive systems, Uncertainty}

\maketitle

\section{Introduction}
\label{sec:Introduction}
Adaptive systems are systems that modify their behavior in response to changes in their environment or in their components. In general, adaptive systems are quite complex to design, verify, and optimize, and therefore simulations are used to analyze both their behavior and their properties of interest. In this context, models are used to represent the relevant behavioral characteristics of the system under study, whereas the simulations represent the evolution of the model over time~\cite{DEVS}. Initially inspired by NASA's original replication approaches~\cite{Grieves2017}, Digital Twin Systems (DTS) have emerged to pair the physical system and its models in order to improve the system's design, operation, and maintenance~\cite{DTConsortiumGlossary,DaliborJRSWWW22}. 

One of the problems of having two systems (the physical one and its digital replica) is that their behavior may not always be consistent. According to Segal's law, ``A man with one watch knows what time it is. A man with two watches is never sure.'' 
To complicate things further, such systems are not free of uncertainty; and in case of inconsistency between the two systems, it is important to understand if this is due to uncertainty or to unexpected divergence. More generally, uncertainty is an inherent property of any physical system, and can be a problem in the particular case of adaptive systems, whose decisions and control mechanisms can be affected by inaccuracies in sensor readings, 
looseness in mechanical parts, or inexact comparisons~\cite{HERZALLAH200582,Calinescu20, RamirezJC12,Esfahani2013,hahner-secrypt,hahner_seams_23}. These uncertainties can lead to erroneous behaviors.

Although in theory simulation models can help mitigate these inaccuracies, and thus correct their affected behaviors, in practice they are not free of uncertainty either. Consequently, we have to deal with both system and model uncertainties, which are of different natures. Sources of uncertainty in the case of models include input accuracy, numerical approximations, method resolution or model fidelity, among others~\cite{Coleman05,Calinescu20}. And just as the uncertainty in physical systems makes their control and operation more difficult~\cite{Mesbah18}, the uncertainty in the simulation models hinders their reliability and the fidelity of their predictions~\cite{dur36061,JV2023}. 

In this paper, we address the problem of having to deal with two versions of the same system (the physical and the digital) that may diverge in their behaviors. Both systems suffer from uncertainty, but we show that this uncertainty can be combined and exploited to improve both systems.
In order to address these issues in the context of DTS, in this work, we propose the following approach: 

\begin{itemize}
\item First, treating the sources of uncertainty, both in the system and in the models, as \textit{first-class citizens}, and representing them using random variables. This way, control systems decisions can be significantly improved. 
\item
The explicit representation of uncertainty will also enable us to quantify it.   
Therefore, we will be able to determine the \emph{reliability} of the simulation of the system, in the sense that we will consider that it is no longer reliable if its uncertainty exceeds a certain threshold.
\item
The quantification of uncertainty will also allow a more accurate comparison of the behaviors of the physical system and the digital twin, determining when the two behaviors are \textit{consistent} or, on the contrary, diverge. 
 In the case of divergent behaviors, detecting them can help trigger any behavior that attempts to investigate the cause (e.g. a broken part, a sensor malfunction or even that the model is no longer reliable) 
  and react accordingly.
In the case of consistent behaviors, we will show how both model and system uncertainties can be significantly reduced by properly combining the individual uncertainties, using sensor fusion techniques~\cite{Elmenreich2002}.
\end{itemize}

The proposed approach will be demonstrated using the incubator case study~\cite{Oakes2023}, a common example of a digital twin of a control system. Despite its tractable size, its subtleties and intrinsic complexities permit illustrating many of the problems of adaptive systems when faced with uncertainty.

Based on our experiments on the case study, we show that treating uncertainty as a first-class citizen enables a deeper understanding of the observed behaviors, leading to an improvement of the control decisions and hence the overall system behavior. 
Although the scope of Digital Twin Systems is quite varied~\cite{BOTTJER2023162}, the incubator serves as a generic exemplar, and we believe our results extend to a wide range of DTS.

The organization of the paper is as follows. 
Section~\ref{sec:Background} describes the context and background of our work, while Section~\ref{sec:RunningExample} 
describes the running example used to illustrate and evaluate our proposal. Then, Section~\ref{sec:Proposal} presents our proposed approach and Section~\ref{sec:Discussion} discusses some of its advantages and 
limitations. 
Finally, 
Section~\ref{sec:Relatedwork} relates our proposal to similar works, and 
Section~\ref{sec:Conclusions} concludes with an outlook on future work.

\section{Background}
\label{sec:Background}
   
\subsection{Uncertainty}

Uncertainty is an inherent property of any system that operates in a real environment or interacts with physical elements or humans. 
Uncertainty is ``the quality or state that involves imperfect and/or unknown information'' 
\cite{JCGM100:2008}. 
Uncertainty can be due to different factors, such as imprecision in the measuring tools; lack of knowledge about the system or its environment; 
incorrect, incomplete, or vague information; unreliable data sources or communication networks; numerical approximations; 
unforeseen, emergent or unpredictable behavior; 
or the inability to determine whether particular events have occurred or not~\cite{JCGM100:2008}.

The purpose of explicitly representing uncertainty is twofold: a software engineer who represents or simulates a system needs to capture the relevant characteristics of uncertainty in a suitable way, while a systems engineer analyses uncertainty to try to remove it, reduce it or mitigate its effects~\cite{2018-moreno-seams-c}. 
Methods to deal with uncertainty in its many forms (objective, subjective, epistemic, aleatory)~\cite{Thunnissen03}, using different approaches such as mathematical and numerical models~\cite{OBERKAMPF2002333}, probabilities~\cite{Finetti2017}, Fuzzy set theory~\cite{Zimmerman01}, variability analysis~\cite{2004-seely-variabilityanalysis-j}, and risk assessment~\cite{2013-rausand-riskassessment-b} have been extensively covered in literature~\cite{TroyaMBV21}.

Uncertainty can be either aleatory or epistemic~\cite{OBERKAMPF2002333}.
\emph{Aleatory uncertainty}
refers to the inherent stochastic variability or randomness of a phenomenon. 
For example, measuring a physical attribute.
This type of uncertainty is irreducible, in that there will always be variability in the underlying variables~\cite{JCGM100:2008}.
\emph{Epistemic uncertainty} 
refers to the lack of knowledge we have about the system or its environment. For example, how the system will be used.
This type of uncertainty is reducible, in that additional information may reduce it. 

We now define a few types of uncertainty that are relevant for our work but we stress that our contribution is not limited to these types of uncertainty.

\emph{Measurement uncertainty} is an aleatory uncertainty that represents the expression of the statistical dispersion of the values attributed to a measured quantity.
Numerical uncertainty comes from operations whose outcome is only approximate. For example, using floating point arithmetic \cite{Goldberg1991},  using numerical solvers to approximate the solution to continuous differential equations \cite{Cellier2006}, or measuring the temperature or length of any physical object, since sensors and measurement instruments are unable to distinguish values below their accuracy.

In the case of numeric values, this type of uncertainty is usually 
expressed in different ways, 
\eg by means of the standard deviation, $\sigma$, of the values of $x$ ($x\pm \sigma$, \eg $3.5\pm 0.01$); using intervals (\eg [$a$..$b$]) according to Uniform or Triangular distributions; or by means of samples~\cite{JCGM100:2008}. 
See~\cite{HackC12} for a survey on this topic. In this study, we followed recommendations from \cite{JCGM100:2008} and expressed the uncertainty by using standard deviation.

Note that uncertainty propagates through the operations performed on variables with uncertainty~\cite{JCGM100:2008}, affecting all related variables. 
Similarly, comparison between uncertain real numbers are no longer Boolean values, but become probabilities~\cite{2020-Bertoa-SoSym-j}. For example, consider the real values $x=2.0$ and $y=2.5$. Using Real arithmetic, $x<y=$ true. However, assuming some given uncertainties, namely $x=2.0\pm 0.3$ and $y=2.5\pm 0.25$, then we obtain that $x<y$ with probability 0.893 \cite{2020-Bertoa-SoSym-j}. In the following, uncertain reals will be {considered} random variables following Normal distributions, otherwise they could be converted to such variables with adjusted standard deviations, as described in the GUM standard~\cite{JCGM100:2008}. They will be compared using \textit{equality in distribution}, \ie two variables are equal if their distributions are the same.  
 
\subsection{Control Systems}

Control system engineering aims to design automated controllers for a process to keep its operating characteristics at a desired \emph{set-point}. To drive the process to the desired step point, a control system use sensors and actuators, respectively, to monitor the state of the process and to modify its state. 

A feedback-based controller utilizes an error signal which is the difference between the actual state of the system and the set-point, to make control decisions. The sensor accuracy directly impacts the quality of the control. Considering the controller of an oven, if the temperature sensor provides a biased value of +5°C compared to the measurand, then the control will set the oven temperature 5°C higher than expected. 

State-of-the-art in control engineering encompasses many levels of control loops. In embedded systems, the control is mainly implemented by using a Proportional-Integral-Derivative (PID) algorithm and its derivatives, since they provide excellent performance and simplicity in tuning. For simpler systems, where hysteresis around the step point are sufficient, the controller algorithms are referred as \emph{bang-bang} or \emph{on/off} controllers. They use the actuator or not, depending on a threshold over and under the desired step point.
See~\cite{KOZAK20141} for an extensive survey on this topic.

\subsection{Digital Twins}

While there is not yet an agreed   definition of  Digital Twin, in this paper we will consider that a \emph{Digital Twin (DT)} is a virtual representation of a real-world entity or process (the \emph{Physical Twin}, PT), synchronized at specific points in time~\cite{DTConsortiumGlossary}. The twinned systems (the DT and PT), the connections between them, and the set of system services compose the so-called \emph{Digital Twin System} (DTS)~\cite{Grieves2017}. 

The services allow exploiting the data exchanged by the two twins in different ways~\cite{TAO2018169,2019-tao-tid-j,Paredis2021}. 
Examples of these services are dashboards to visualize and display data; Machine Learning components to support decision-making processes and predict changes in the physical twin over time in order to accomplish, \eg preventive maintenance;  monitors to detect anomalies and trigger alerts to users; or algorithms to improve the system performance, conduct fault diagnosis or what-if analysis. 
A DT can also have the capability of modifying the structure or parameters of its corresponding PT based on observations in use-cases like self-adaptation, self-learning, or self-reconfiguration. 
\section{Running Example - The Incubator}
\label{sec:RunningExample}
\begin{wrapfigure}{r}{0.42\textwidth}
    \centering
    \vspace{-6mm}
    \includegraphics[width=0.4\textwidth]{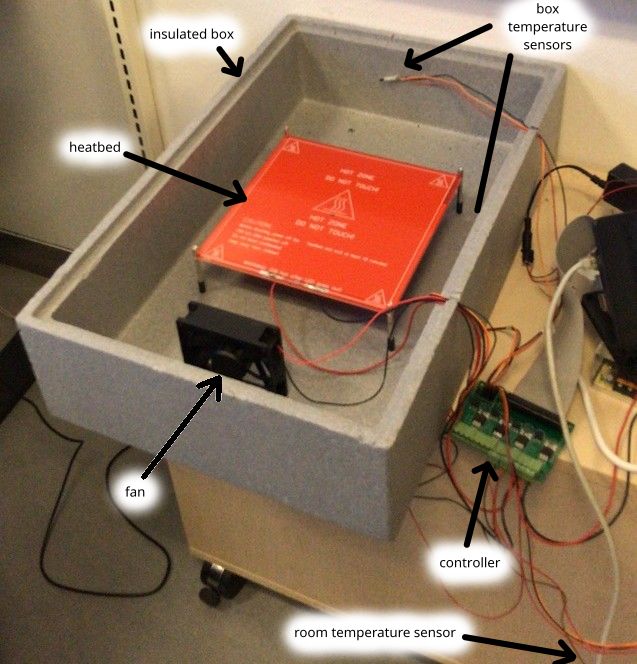}
  \caption{The experimental incubator (lid open). }
  \label{fig:incubatorPhoto}
\end{wrapfigure}

The incubator consists of an insulated box with a heating element and a fan. The controller can read values from two temperature sensors inside the box, and actuate on a fan and the heat bed by switching them \textit{on} or \textit{off} (see Figure \ref{fig:incubatorPhoto}). 
The controller's goal is to maintain a stable temperature for objects inside the insulated box~\cite{Feng2021a,Feng2022a}.

\begin{figure*}[t]
  \centering
  \includegraphics[width=\linewidth]{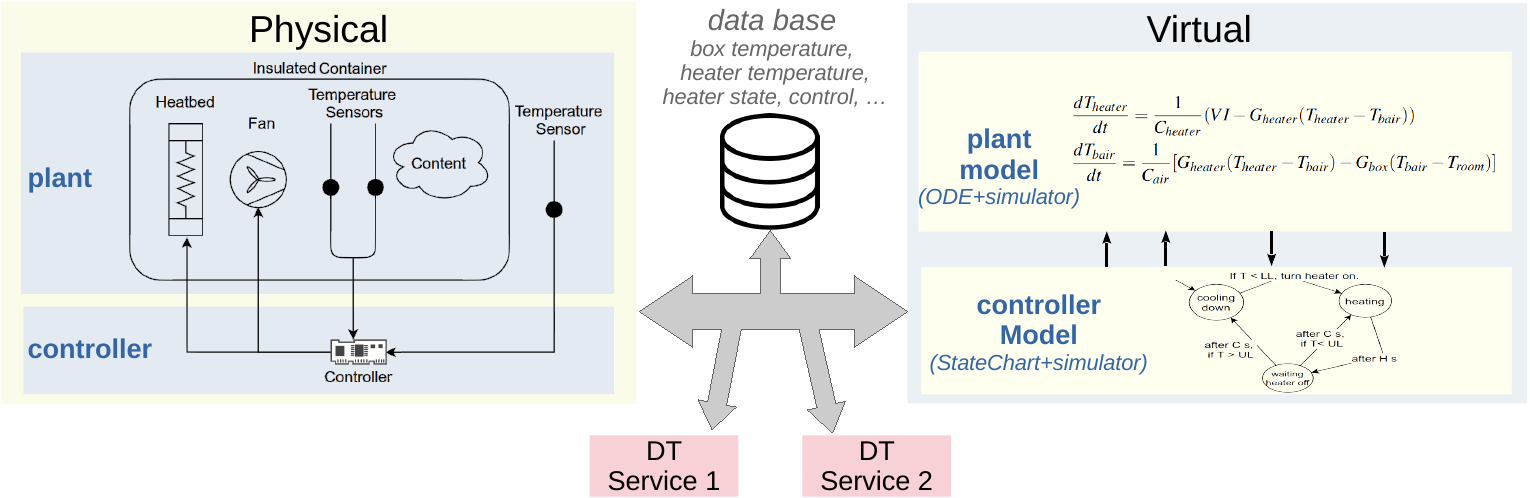}
  \caption{A schematic overview of the incubator digital twin setup.}
  \label{fig:incubatorDtSetup}
\end{figure*}

A conceptual diagram of the incubator Digital Twin System is represented in Figure~\ref{fig:incubatorDtSetup}. The left-hand side represents elements of the Physical Twin, shown in Figure~\ref{fig:incubatorPhoto}, with a separation between the plant and the controller. The implemented controller is a \emph{bang-bang} controller. On the right-hand side, the digital Twin, aimed to mimic the physical twin, is represented. The plant model is specified by ordinary differential equations and can be simulated (by using, \eg a Euler solver). The controller is modeled by a State Chart and can be simulated too. The data exchanged between the controller and the plant is the same in the physical and the digital twins. Typically, both twins write measured/computed data in the database and can consume reconfiguration commands. 
A set of services completes the architecture. The services also read and write from and to the database. Examples of such services are presented later in this paper. 

\subsection{Physical Twin Uncertainty}
\label{sec:PTRunningExample}

The actual physical incubator 
uses two DS18S20 sensors\footnote{\url{https://www.analog.com/media/en/technical-documentation/data-sheets/DS18S20.pdf}} to sense the temperature inside the box, and one to sense the room temperature.\footnote{Note that so far the incubator has been tested under the hypothesis of constant room temperature.}
From their datasheets, these sensors have an accuracy of $\pm 0.5^{\circ}C$ and they can be read only once every two seconds. The controller is hosted by a Raspberry Pi. The controller runs every three seconds, and it averages the temperature values given by {the two box} sensors to obtain the internal temperature of the box.  
We consider the accuracy depicted in the datasheet follows a uniform probability density function, so that the standard uncertainty $\sigma_{1}$ and $\sigma_{2}$ of respective sensors $S1$ and $S2$ can be approximated to $0.5/\sqrt{3}= 0.289$. We used this value when making the uncertainty of the sensor explicit. We are aware that other sources of uncertainty in the physical system may also exist, like the time between the actual measure of the value by the sensor and its use by the controller; however, to keep the example focused, 
we only {considered} the uncertainty from the sensors. 

   \begin{figure}
    \centering
    \includegraphics[width=\columnwidth]{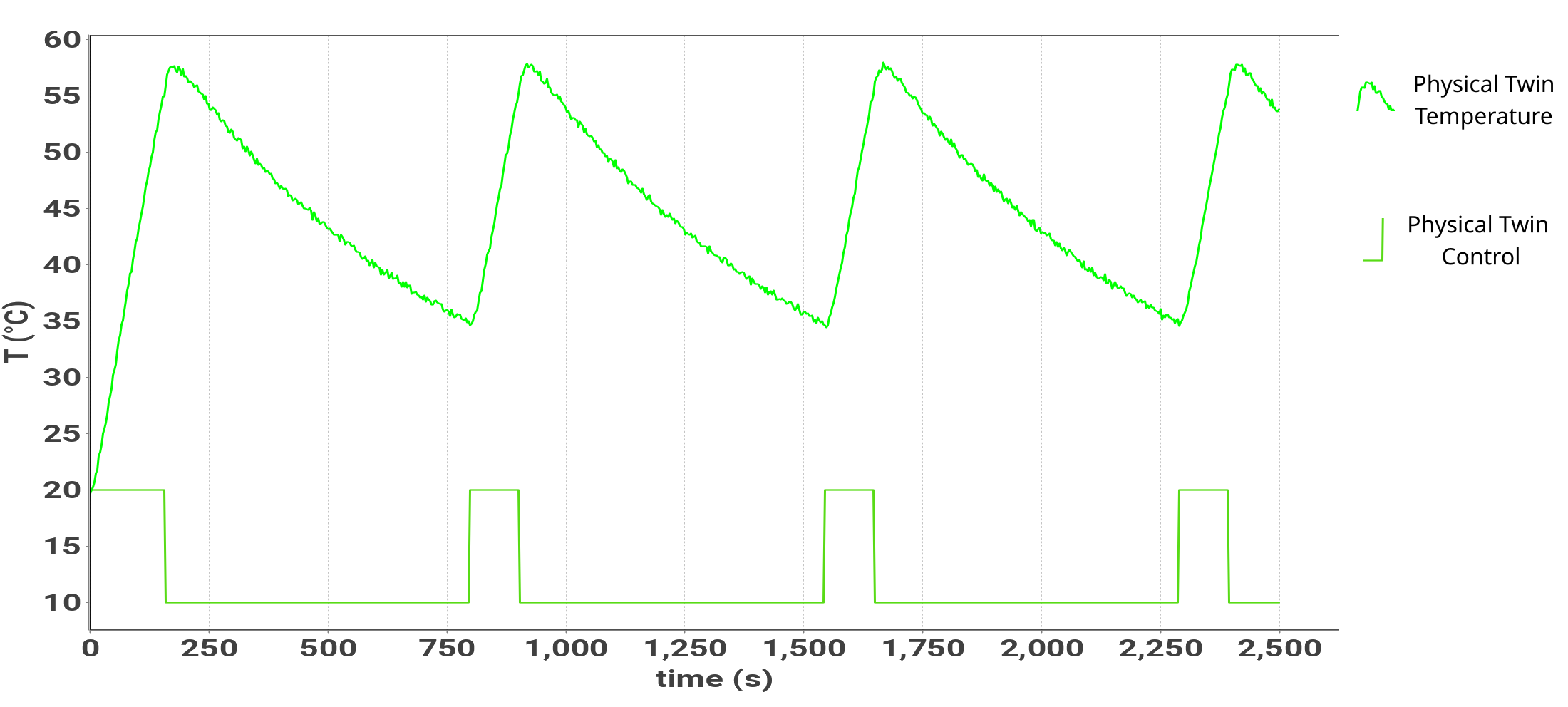}
    \caption{Evolution of the incubator box temperature and the heater control of the classical control (Physical Twin).  The non smoothness of the temperature curve highlights the presence of stable white noise in the incubator temperature.} 
    \label{fig:physical_noisy_run}
  \end{figure}

Two issues 
are worth noting at this point.
First, we assume that the uncertainty from sensors remains constant since the noise between successive sensings of the measurand are not correlated random variables.  Second, when using the actual sensors, we obtain noisy variables as represented in Figure~\ref{fig:physical_noisy_run}. 

\begin{figure}
    \centering
    \includegraphics[width=\columnwidth]{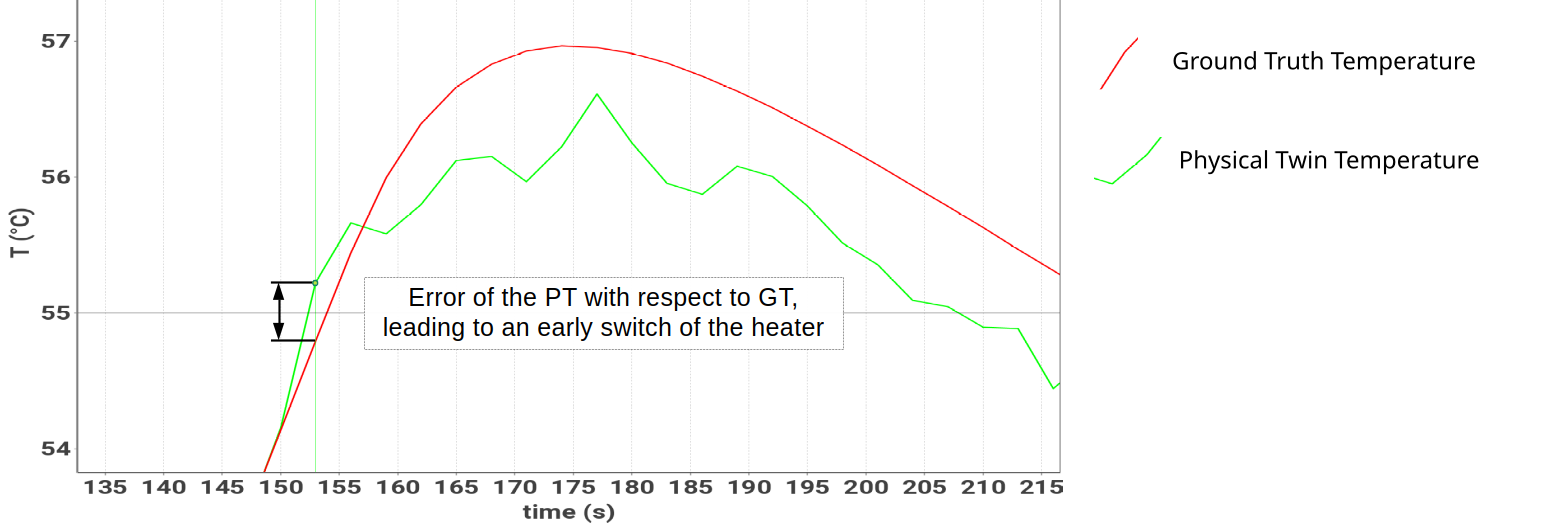}
    \vspace{-4mm}
    \caption{Error of physical twin with respect to measurand system (ground truth), leading to an early switch of the heater.}
    \label{fig:zoomError}
  \end{figure}

It is important to note that since we are using a bang-bang controller, the time at which the control starts and stops the heater is not always the same, depending on the noise or the considered uncertainty. This may cause different executions to {deviate} in time from each other, see Figure~\ref{fig:zoomError}.  
This deviation will be characterized in Section~\ref{sec:Discussion}.

In the experiments conducted in this paper, we had to compare with the ground truth, which is never perfectly obtained by measurements. To enable such comparisons, we used a synthetic measurand defined by differential equations. We also used a synthetic physical environment where we replicated the noises introduced by the sensors according to their specifications.

\subsection{Digital Twin Uncertainty}

There are different models of the incubator, which take into account more or less influential parameters. In this study, we consider a model based on ordinary differential equations according to Newton's law of cooling. The considered parameters are the thermal conductivity of the air inside the box and of the heater; the heat transfer capacity of the box and the heat transfer capacity of the heater; and the voltage and current in input of the heater when \textit{on}. All these parameters were calibrated to match as closely as possible the observations of the actual incubator. However, they are all subject to uncertainty. For example, the heat transfer capacity is calculated from the materials and the mass of the heater. 
These two values are obtained using heater data sheet knowledge and instruments, which both introduce uncertainty. 
We model them using standard deviations to allow taking these measurement uncertainties into account during the simulations.

\begin{figure}
    \centering
    \includegraphics[width=.9\columnwidth]{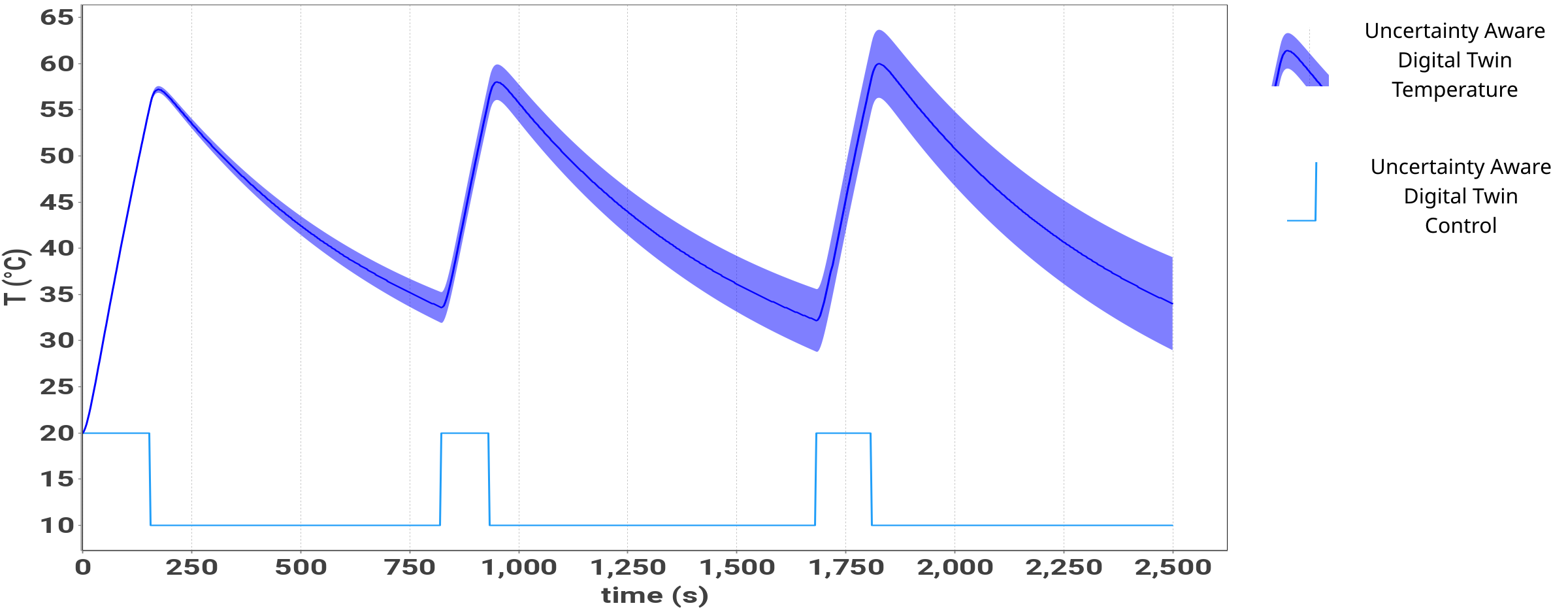}
    \caption{evolution of the incubator temperature and the heater control of the uncertainty-aware model based control (Uncertainty-Aware Digital Twin). The uncertainty is represented around the nominal temperature value.  It highlights the increase of uncertainty along the simulation.}
    \label{fig:model_uncertain_run}
  \end{figure}

Additionally, the model is approximated by a solver, which in our case uses a forward Euler integration method. The discretization performed by the solver introduces numerical approximations and consequently uncertainties. The Euler method is a first-order method, which means that the local error (error per step) is proportional to the square of the step size; and so is the related uncertainty.
Contrary to sensor uncertainty, model uncertainty increases with time as input uncertainties propagate through the model and equations are approximated by the solver. As a result, the uncertainty grows along the simulation---see Figure \ref{fig:model_uncertain_run}.

\section{Proposal}
\label{sec:Proposal}
The previous section has presented the two main artifacts of our DTS, the physical and the digital twins. Both are subject to uncertainty. The uncertainty of the Physical Twin comes from sources such as the sensors' accuracy or discretization period. The uncertainty of the Digital Twin comes from numerical approximations, parameter approximations and from the model itself. Thus, the uncertainties of both twins come from different and independent sources. 
In this section, we explain how making the uncertainty in each artifact explicit can be used to improve the overall behavior of the DTS.

\subsection{Considering Uncertainty as a First-class Citizen}

We propose to explicitly model the various sources of uncertainty in the DTS. The main idea consists in assessing systematically the sources of uncertainty to enable the conscious use of all the data in a DTS. This means that random variables must be leveraged in both the PT, the DT and the associated services. 

Following the ISO Guide to Measurement Uncertainty~\cite{JCGM100:2008}, the uncertainty in each of the values can be represented by a Gaussian distribution, whose uncertainty is given by the standard deviation of the distribution. Then, we can consider that an execution trace (of either the system or the model) is no longer a set of values $Y=\{y_i\}_{i=1}^n$ but a set of random variables, where each random variable is defined by both an actual value and its standard deviation:  $Y=\{y_i \pm \sigma_i\}_{i=1}^n$.

This is intuitively depicted in Figure~\ref{fig:model_uncertain_run}, which shows not only the curve with the nominal incubator temperature but also its uncertainty. We can see how the temperature evolution is now represented by what is called a ``flow pipe,'' which wraps around the initial curve, enveloping it; and showing the uncertainty.

Considering the use of uncertainty in the PT, we used the knowledge about the sensors' accuracy to define the box and room temperatures as random variables with a standard deviation of 0.289 (see Section~\ref{sec:PTRunningExample})\footnote{These adaptations can be done directly in the implementation of the sensor acquisition, or offered as a service of the DTS, which could create random variables from the raw sensor readings in real-time.}.

We explained that the controller of the PT averages the values from two sensors to obtain the box temperature, each of which has a standard deviation of 0.289. In this case, the uncertainty decreases since we have two independent sources of evidence for the same value. The resulting uncertainty $\sigma_T$ of the temperature used by the controller is 
$\frac{1}{2}\sqrt{\sigma_{1}^2+\sigma_{2}^2}\approx 0.204$
which is less than $0.289$. 
This example is a good illustration of our first proposal.  
Sensors are noisy, and it is well-known that averaging independent sensor values reduces the effect of the noise~\cite{Elmenreich2002}.
However, it is not always obvious what is the resulting uncertainty after the averaging. We advocate that making uncertainty explicit helps in the further operation with its values. 
Here, since the box temperature is a random variable, the controller may not compare whether the temperature is above or below a specific threshold. Instead, it may compare whether the confidence of being above or below is greater than a specific confidence level (typically 95\%). In this case, using random variables is a natural enabler for a systematic and lightweight form of stochastic control. In our experiments, such modifications provided a control closer to the ground truth than the classical system control that does not consider uncertainty.

\begin{figure}
    \centering
    \includegraphics[width=\columnwidth]{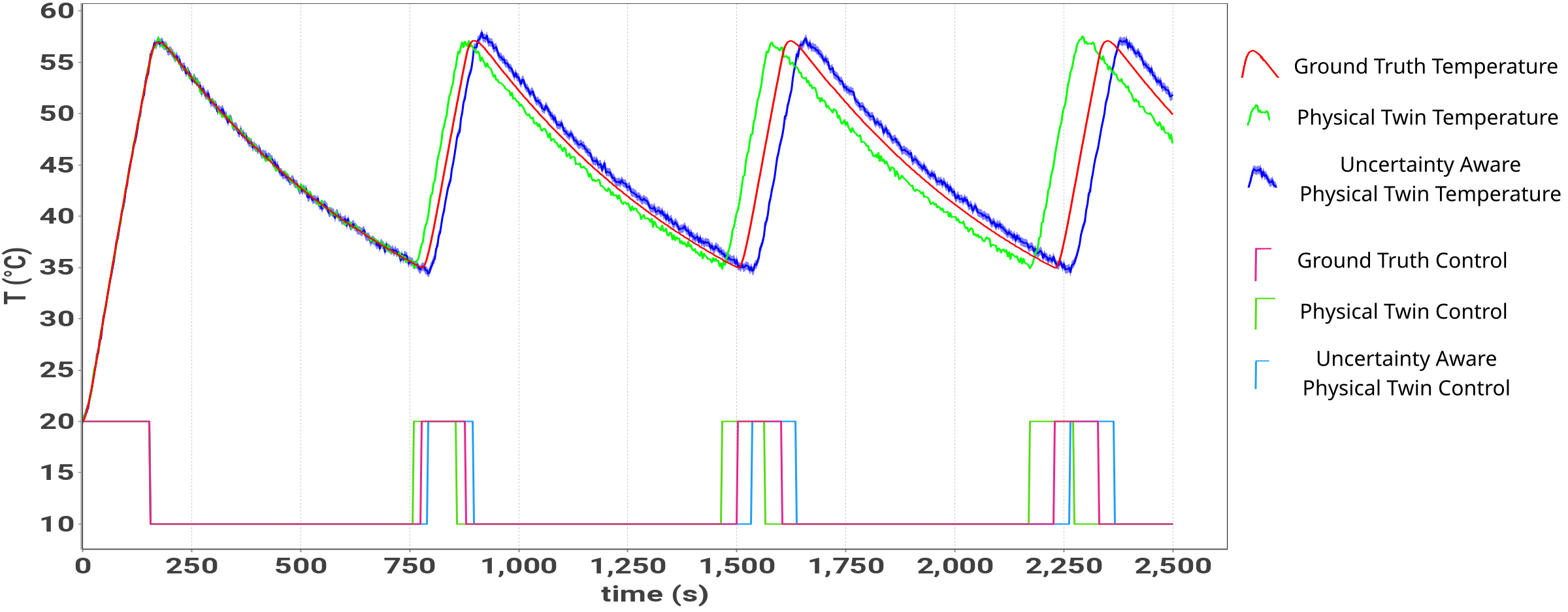}
    \caption{ Comparison of the evolution of the incubator temperature and the heater control between the measurand \textit{system}  (Ground Truth), the classical control \textit{system} (Physical Twin) and the uncertainty-aware control (Uncertainty-Aware Physical Twin). Confidence = 95\%.}
    \label{fig:uncertaintyAwareControl}
  \end{figure}

This is illustrated in Figure~\ref{fig:uncertaintyAwareControl}, where the \textit{red} time series represents the ground truth (\ie the measurand), the \textit{green} time series the classical control system (\ie the PT) and the \textit{blue} time series the control with uncertainty as a first-class citizen (\ie the UAPT). We can see how the blue line is closer to the red one, reacting to the temperature thresholds more accurately. We will discuss this in more detail in Section~\ref{sec:discussion1}.

\begin{figure}
    \centering
    \includegraphics[width=\columnwidth]{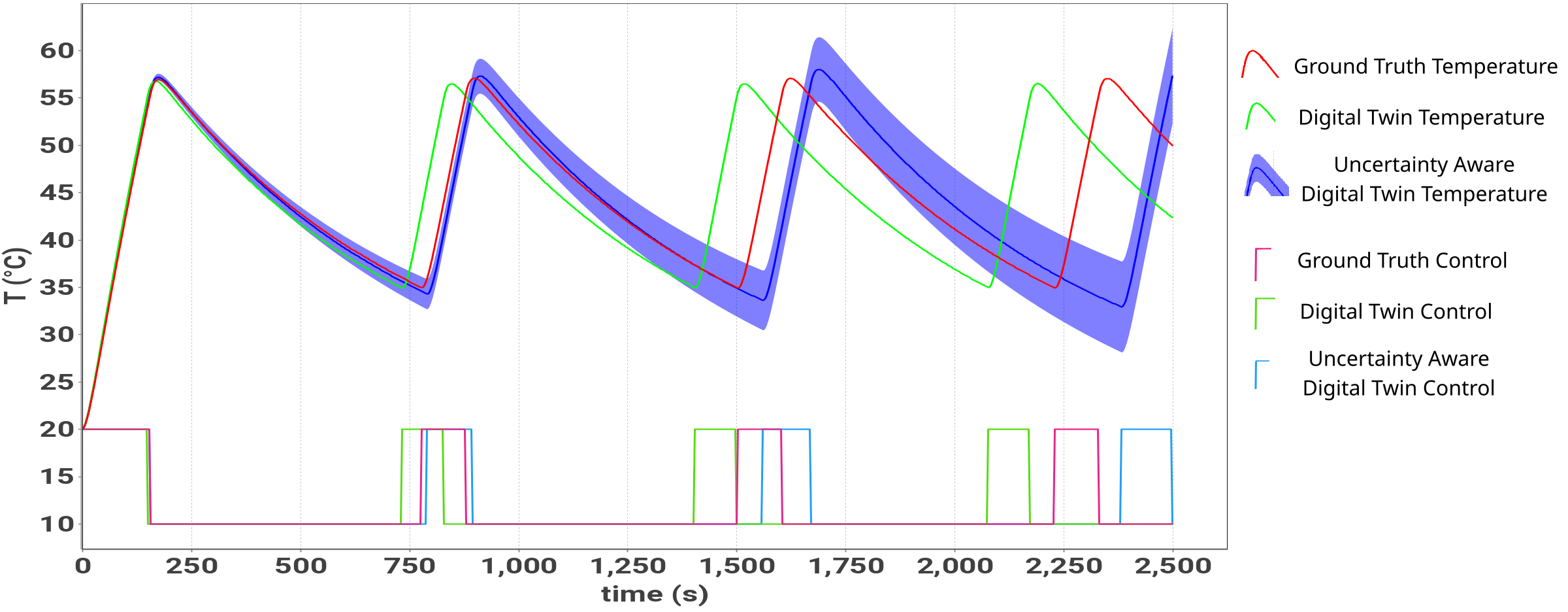}
    \caption{Comparison of the evolution of the incubator temperature and the heater control between the measurand \textit{system}  (Ground Truth), the model-based control \textit{system} (Digital Twin) and the uncertainty-aware model-based control (Uncertainty-Aware Digital Twin). Confidence = 95\%.}
    \label{fig:uncertaintyAwareModels}
  \end{figure}

Considering the use of uncertainty in the DT (\ie the simulation models), all the arithmetic behind the ODE and the solver should make use of random variables to understand how the uncertainty from the random variables used in the parameters and the inputs propagates to the outputs. Additionally, it {motivates} us to characterize the numerical error introduced by the equation solver. 
It is not always obvious to figure out how fast the uncertainty grows throughout the simulation, depending on the size of the discretization time step or the change of uncertainty in the sensors.
By making uncertainty explicit, users are faced with the fact that computing the evolution of the box temperature from noisy sensors with an Euler solver and a discretization time step of 1ms, for 2500 seconds, leads to a box temperature with a standard deviation of 2.52 °C, \ie with an error of $\pm$5.04 °C (at 95\% of confidence, see Figure~\ref{fig:uncertaintyAwareModels}). It is up to the user to consider this information relevant or not and to handle it accordingly. However, we believe it should not be ignored. 

To make the above consideration explicit, we propose to define the notion of \textit{reliability} of the simulation based on the quantification of uncertainty. In other words, it is important, when a simulation is used in a specific context, to clearly define the conditions under which the simulation is considered sufficiently meaningful (and reliable). 
The amount of acceptable uncertainty depends on various factors, such as the characteristics of the modeled physical phenomena or the robustness of the controller. However, we advocate for an explicit definition of the acceptable amount of uncertainty on each output random variable, above which the simulation is not considered anymore as a model of the reality (in the sense of \cite{minsky1965matter}), \ie the model cannot be used to confidently answer questions about the corresponding reality. For instance, for the incubator example, an error of $\pm$5°C may be tolerated if the goal of the simulation is to dimension the incubator power supply; however, it is not acceptable if the goal is to use the incubator to hatch eggs\footnote{Within a range of 35 to 40.5°C there is the possibility that eggs will hatch. The optimum (for hens) is 37.5 °C. Above this temperature, in addition to reduced hatching, there will be an increase in the number of crippled and deformed chicks. Above 40.5°C no embryos will survive. \url{https://brinsea.co.uk/latest/wp-content/uploads/2017/04/What-if-the-power-goes-off-2010.pdf}}.

Technically speaking, the definition and handling of such random variables as first-class citizens ultimately rely on technical libraries and require changes both in the physical and digital twins of the system. In the modeling and simulation context imposed by a DTS, a suitable library should (1) allow the usual arithmetic operations imposed by classical algorithmic and ODE solving, (2) allow for probabilistic comparison of random variables as required by stochastic process control, and (3) use closed-form solutions to avoid too much overhead, {which would make} real implementations unrealistic. In this paper, we used the Java Library described in~\cite{2020-Bertoa-SoSym-j}, which fits such requirements.

To distinguish the different systems and models we are going to compare, we will use identifiers for them:

\begin{itemize}
\item \emph{Ground Truth (GT)}: The measurand system, which corresponds to the system behavior with ``perfect'' sensors, \ie uncertainty $=0.0$.

\item \emph{Physical Twin (PT)}:  Implementation of the system in the Raspberry Pi. The uncertainty from the sensors is not considered by the controller.

\item \emph{Uncertainty-aware PT (UAPT)}:   same as the PT, but the controller uses random variables (uncertain Reals and Booleans) in its decisions.

\item \emph{Uncertainty-aware DT (UADT)}: Simulation model that uses random variables in both the plant and the controller.
\end{itemize}

\subsection{Uncertainty Mitigation By Using the DTS}
\label{sec:uncertaintyMitigation}

We explained in the previous subsection that both the PT and the DT have uncertainties. 
In this paper, we take advantage of the fact that we have two synchronized systems executing the same behavior with different sources of uncertainty and therefore representing two independent sources of information. As we shall see, this will enable the mitigation of uncertainty. 
We rely on the extensive work about forecast prediction, where since the late 1960s it is well known that if you take several forecasts and average them, then the resulting aggregated forecast outperforms the individual forecasts~\cite{CLEMEN1989559}. The most accurate way to average them is to use a weighted average, in which each forecast is weighted by its inverse ``error''~\cite{Winkler83}; this is in essence what the Kalman filter does~\cite{Bishop2001}. In our context, that error corresponds to their uncertainty. This way, less weight is given to the least certain predictions. 

In our case, the PT has uncertainty in the box temperature $T_{p}$ due to its implementation characteristics; and the DT has uncertainty in the box temperature $T_{d}$ due to modeling characteristics. This means that $T_{p}$ and $T_{d}$ can be modeled by two independent variables that follow Normal distributions with standard deviations $\sigma_p$ and $\sigma_d$, respectively. 
Their \textit{weighted average} is then given by another Normal distribution, $T_{A}=\frac{1}{(\sigma_{p}^2 + \sigma_{d}^2)}(\sigma_{d}^2 {T_{p}} + \sigma_{p}^2 {T_d})$, 
whose variance is $\sigma_A^2=(\sigma_{p}^2\sigma_{d}^2)/(\sigma_{p}^2+\sigma_{d}^2)$. 

Since for any two positive numbers $a,b$ we know that $ab/(a+b)\leq min\{a,b\}$, we conclude that the uncertainty of the weighted average of two uncertain numbers will always be less than the uncertainty of any of the two.

   \begin{figure}
    \centering
    \includegraphics[width=0.5\columnwidth]{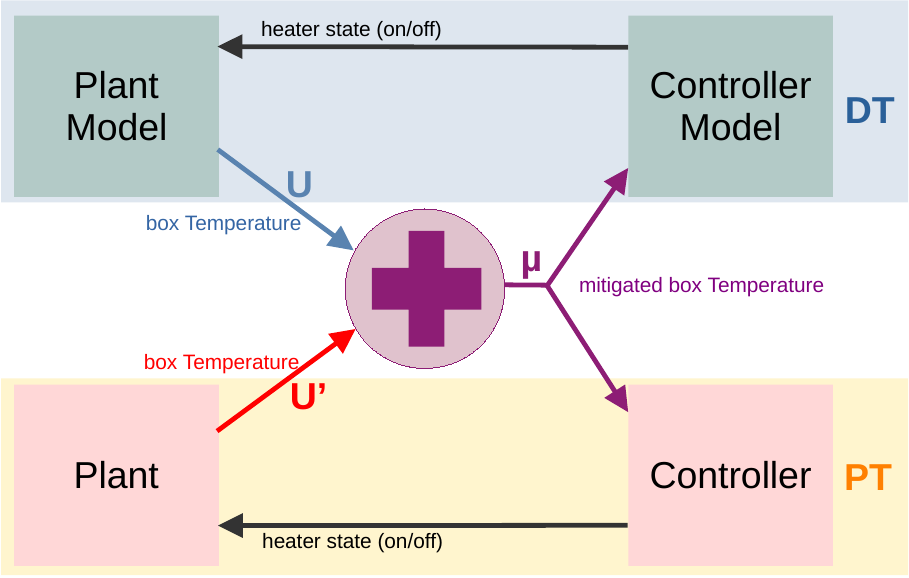}
    \vspace{-2mm}
    \caption{Architecture to Mitigate Uncertainty ($\mu$) in a DTS context (MDTS).}
    \label{fig:mitigated_uncertainty_by_dt}
  \end{figure}

\begin{figure}
    \centering
    \includegraphics[width=.9\columnwidth]{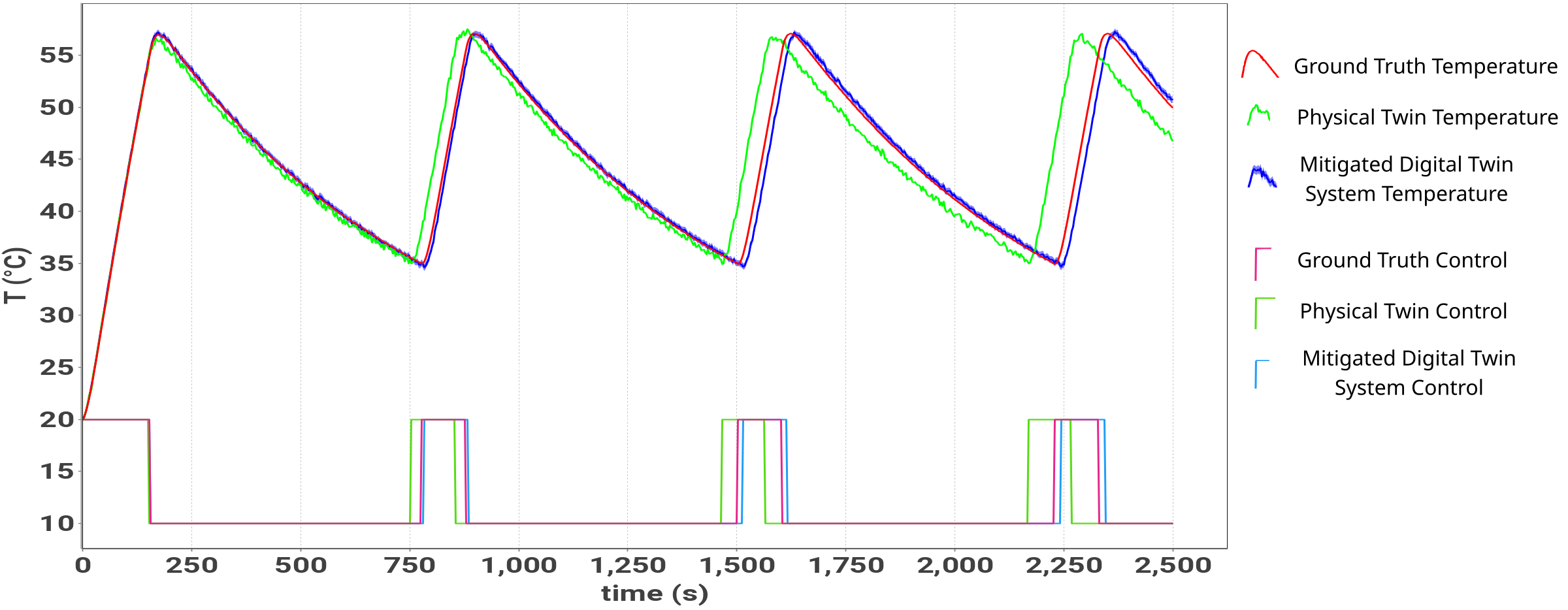}
    \caption{Comparison of the evolution of the incubator temperature and the heater control between the measurand \textit{system}  (ground truth), the classical control \textit{system} (Physical twin) and the mitigated control \textit{system} (mitigated digital twin system). Confidence = 95\%.}
    \label{fig:mitigatedComparison}
  \end{figure}

To realize this averaging, we propose a digital twin service that implements the architecture described in Figure~\ref{fig:mitigated_uncertainty_by_dt}. The box temperatures provided by both twins are averaged before being provided to the controllers of each twin. As explained, the mitigated uncertainty $\mu$ from Figure~\ref{fig:mitigated_uncertainty_by_dt} is smaller than the minimum between the uncertainties $U$ and $U'$ of the temperatures from the plant and from its model. Since the input uncertainty $\mu$ is now lower in both controllers, they can perform a more accurate control of the box temperature. This is illustrated in Figure~\ref{fig:mitigatedComparison}, where the behavior of the controller with the mitigated uncertainty ($\mu$) is much closer to the intended ground truth and consequently drifts at a slower rate from it. 

We refer to the uncertainty-aware DTS making use of mitigation as the \emph{Mitigated Digital Twin System (MDTS)}. It uses the UADT and UAPT, but their controllers use the averaged uncertainty $\mu$. 
By using this method, we obtain a better control since controllers are fed with a less uncertain box temperature value. However, the uncertainty of the plant model continues to grow throughout the simulation. To avoid this phenomenon, we will use the notion of\textit{ model reliability}. Thus, when the model simulation is too close to its reliability limit, we reset its box temperature with a more certain value, \ie the mitigated one; reducing the box temperature from $U$ to $\mu$ and consequently taking away the plant model from its reliability limit. 
{The same kind of approach is used in the Kalman filter~\cite{Mezic2004}, but the resetting is done at each simulation step.} 

To avoid too many synchronizations between the PT and the DT, for instance to be used in a distributed DTS setup, we realize the mitigation only when required, \ie when the model simulation gets too close to its reliability limit. 
Again, we take advantage of the explicit representation of uncertainty and the possibility to consciously define the simulation reliability, to (1) avoid the increasing uncertainty in the plant model throughout the simulation, and (2) limit the number of re-settings to their minimum to maintain the reliability of the model simulation.

Note that this averaging makes sense only if applied to values that are \textit{consistent}, \ie when different sources of information represent the same reality. 
To ensure that the averages yield correct results, we defined a notion of consistency between random variables and used it to detect divergences between the PT and DT.

\subsection{Defining Consistency Between Values Represented by Random Variables}

In the previous subsection, we explained how PT and DT uncertainties can be averaged to mitigate the overall uncertainty and, consequently, improve the behavior of the DTS. However, averaging data if conditions change in the physical environment can be dangerous as the results may become meaningless or, worse, hide the change---which could have catastrophic consequences. For instance, if someone opens the lid during incubation or the box breaks, averaging the inconsistent values may mask the actual change in the physical environment, and potentially lead to a fire.

We exploit our uncertainty quantification to define the \textit{consistency} (denoted as $\Bumpeq$) between two random variables. For this, suppose two values $v_{p}$ and $v_{d}$ that represent the same physical value at time $t$ as provided by the PT and the DT. Each value has an explicit uncertainty, $\sigma_{p}$ and $\sigma_{d}$. Then, they define two random variables $V_p$ and $V_d$ with Normal distributions whose means are the values $v_{p}$ and $v_{d}$, and their standard deviations are the respective uncertainties. We say that both values are \textit{consistent} with a given confidence level of $c$, e.g., 95\%, if 
the intervals $[v_{p}-K\sigma_{p},v_{p}+K\sigma_{p}]$ and $[v_{d}-K\sigma_{d},v_{d}+K\sigma_{d}]$ overlap; \ie
\begin{equation*}
    [v_{p}-K\sigma_{p},v_{p}+K\sigma_{p}] \cap [v_{d}-K\sigma_{d},v_{d}+K\sigma_{d}] \neq \emptyset
\end{equation*}

In this equation, $K$ is a real number that represents the extended uncertainty needed to ensure that the percentage of values that lie within an interval estimate of $[x-K\sigma,x+K\sigma]$  from a Normal distribution $N(x,\sigma)$ is at least $c$. For example, $K=1$ for $c=0.68$, $K=2$ for $c=0.95$, and $K=3$ for $c=0.997$. In statistics, this is known as the 68–95–99.7 rule. In the following, we will consider that $c=0.95$, and therefore we will take $K=2$. Note that this value $K=2$ corresponds to what is called \textit{extended uncertainty} in the GUM~\cite{JCGM100:2008}.

This definition of consistency between values that are represented by random variables ($\Bumpeq$) can be extended to return not only a Boolean value that expresses whether two variables are consistent or not, but a degree of consistency, expressed using a real number between 0 and 1.  
First, if one of the 95\% intervals (\ie $\pm 2\sigma$) of one of the variables is fully contained in the other, then the degree of consistency is 1. If the two intervals do not overlap, the degree of consistency is 0. Otherwise, if the intervals overlap, the degree of consistency is simply defined by the ratio between the intersection of the $\pm 2\sigma$ intervals and the union of $\pm 2\sigma$ intervals; \ie 

\vspace{-3mm}
\begin{small}
\begin{equation*}
V_p\Bumpeq V_d = \frac{min\{v_{p}+2\sigma_{p}, v_{d}+2\sigma_{d}\} - max\{v_{p}-2\sigma_{p}, v_{d}-2\sigma_{d}\}}{max\{v_{p}+2\sigma_{p}, v_{d}+2\sigma_{d}\} - min\{v_{p}-2\sigma_{p}, v_{d}-2\sigma_{d}\}} 
\label{eq:2}
\end{equation*}
\end{small}

The less the intervals of uncertainty of the two variables overlap, the closer to zero the degree of consistency is.

Note that this consistency operator is very different from the simulation's reliability one. Operator ``$\Bumpeq$'' does not represent the confidence we can have in the data, which may have a lot of uncertainty. Instead, it represents the likelihood that the two values belong to \emph{close enough} distributions. 
Note also that this is different from the probability of the two variables being \emph{equal in distribution}~\cite{2020-Bertoa-SoSym-j}, and also different from classical statistical distance functions between random variables~\cite{Rachev2013}.

\subsubsection{Consistent Behaviors}\label{sec:consistentBehaviors}

The consistency operator ($\Bumpeq$) was defined between values that are represented by means of random variables.  This operator can also be extended to uncertain behaviors, \ie sequences of uncertain values that correspond to the execution traces of either the DT or the PT. 

Given two uncertain behaviors $X=\{x(t)\pm\sigma_x(t)\}_{t=1}^n$ and $Y=\{y(t)\pm\sigma_y(t)\}_{t=1}^n$, we say that, given a confidence level $c$, \eg $c=0.95$,  $X$ and $Y$ are consistent behaviors (noted as $X \Bumpeq Y$) iff
    $x(t) \Bumpeq y(t) \geq c, \forall t=1..n$.

Note that one could also define weaker versions of this operator. For example, we can request that  $x(t) \Bumpeq y(t) \geq c$ for only a percentage of the values of $t$, and not for all of them.
In other words, trace consistency can be adapted to different types of more or less smooth processes, but it always defines when two traces should be considered to represent the same measurement behavior.

\subsubsection{Inconsistent Behaviors}\label{sec:inconsistentBehaviors}

In this paper, we are also interested in detecting inconsistent behaviors. More precisely, we are interested in detecting when two consistent behaviors start diverging and become \textit{inconsistent}. 

An inconsistency occurs at one moment in time (\eg at $t_0$) when  $x(t_0)\Bumpeq y(t_0)\leq R$, being $R$ a threshold that represents the minimum degree of consistency below which the two values cannot be considered consistent. For example, we can set $R=0.05$. Note that, by definition, if we set the value of $R$ to $0$, we request that the two uncertainty intervals be disjoint. 
Intuitively, this means that the flow pipes enveloping the two paths do not overlap, or overlap so little that they cannot be considered similar. 

From our experiments, it is important to check the consistency of the behaviors before averaging their uncertainties to avoid hiding a divergence between the twins. The detection and potential reactions to inconsistent and divergent behaviors are discussed later in Section~\ref{sec:managingInconsistentBehaviors}.

\section{Discussion}
\label{sec:Discussion}

In the previous section, we have defined how to capture model and physical uncertainties in the variables by means of random variables that are able to quantify and operate with their uncertainty. We have also described how to mitigate the uncertainty of the inputs of the controllers by combining the model and system uncertainties using their weighted average. Finally, we have defined and characterized what we understand by consistent behaviors and how to detect inconsistencies that represent a divergence in the behaviors of the twins.

In this section, we discuss in more detail the improvements that can be achieved with these definitions by studying a large number of experiments based on the incubator system.

\subsection{Improving the Behavior of the System with Uncertainty-aware Controllers}\label{sec:discussion1}

Our goal was to create a control system that is more accurate to the ground truth by taking into account the uncertainty of the system, as compared to the classical approach where uncertainty is ignored. To achieve this, we used three models represented by red, green, and blue lines. The red line represents ground truth (GT), the green line represents the classical model (PT), and the blue line represents our proposal (MDTS) that considers uncertainty and its mitigation (see Figure~\ref{fig:mitigatedComparison}).

\begin{figure}
    \centering
    \includegraphics[width=.7\columnwidth]{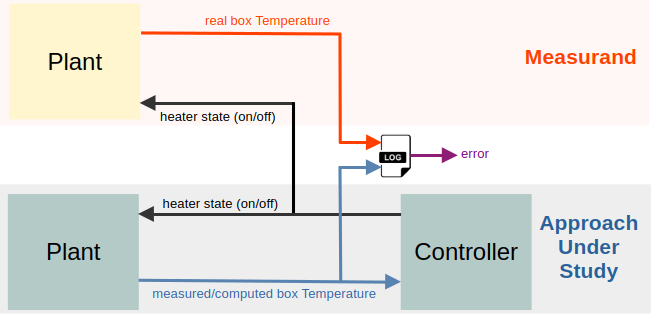}
    \caption{Experimental setup to measure the error of an approach under study with respect to the measurand}
    \label{fig:architectureForErrorMeasurement}
\end{figure}

\begin{figure}
    \centering
    \includegraphics[width=.7\columnwidth]{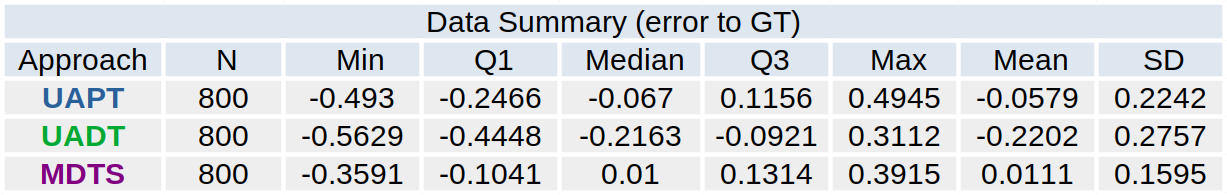}
    \includegraphics[width=.7\columnwidth]{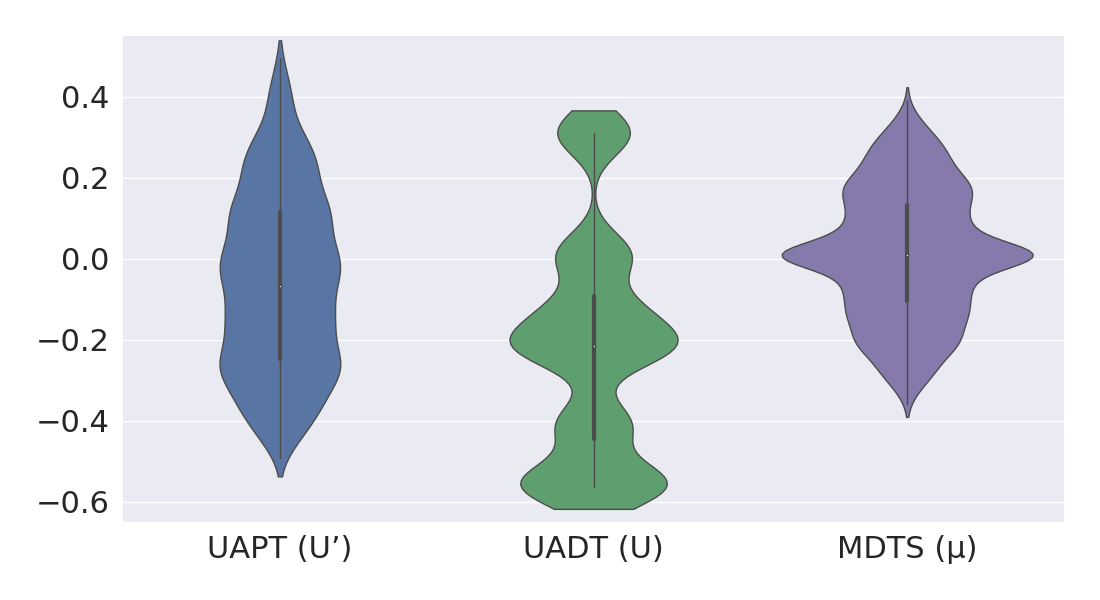}
    \vspace{-2mm}
    \caption{Error with respect to measurand at the heater switch time for different configurations, along 100 runs of 2500 seconds each.}
    \label{fig:box-plot_error_to_measurand}
  \end{figure}

Previous figures \ref{fig:uncertaintyAwareControl}, \ref{fig:uncertaintyAwareModels} and \ref{fig:mitigatedComparison} were obtained by single executions of the system and are therefore subject to change depending on the noise of each run. They provide a simple way to visualize whether a specific approach is ``closer'' or not to what would be a control system based on the box temperature from the measurand. The deviation in time between the different approaches comes from the use of an erroneous box temperature by the controller, compared to the actual one in reality (see Figure~\ref{fig:zoomError}). 
In order to quantify this error, we set up experiments where the measurand is run conjointly with the approach under study. More precisely, the measurand of the plant is controlled by the controller of the approach under study. This experiment allows comparing what actually happens in reality with what is perceived by the controller (see Figure~\ref{fig:architectureForErrorMeasurement}).  
Based on the architecture depicted in Figure~\ref{fig:architectureForErrorMeasurement}, we logged 100 times executions of a 2500-second incubation, and measured the difference between the perceived temperature and the actual one each time the controller decided to switch the heater \textit{on} or \textit{off}. We measured the error of 3 different approaches: the uncertainty-aware physical twin approach (UAPT), the uncertainty-aware digital twin (UADT), and the mitigated approach (MDTS) proposed in Section~\ref{sec:uncertaintyMitigation}. 
The results are presented in Figure~\ref{fig:box-plot_error_to_measurand}, where we compare the errors from the Uncertainty-Aware Physical twin (UAPT), the Uncertainty-Aware Digital Twin (UADT) and the Mitigated Digital Twin System (MDTS), using a violin plot~\cite{violinplots}. These violin plots combine a box plot with a kernel density plot, providing a compact representation of the distribution and density of the errors for each approach. The results represent a significant reduction of the error in the MDTS compared to both the UAPT and the UADT, confirming the benefits of the proposed approach.

\begin{figure}
    \centering
    \includegraphics[width=.7\columnwidth]{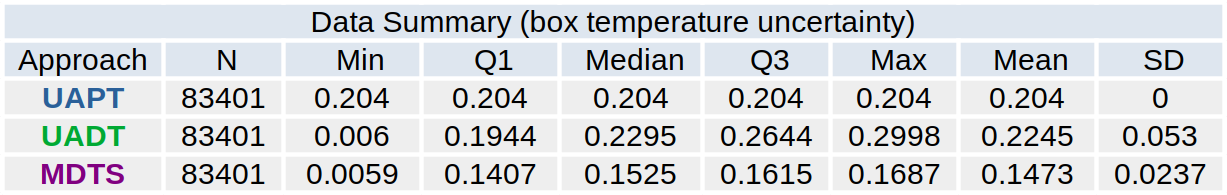}
    \includegraphics[width=.7\columnwidth]{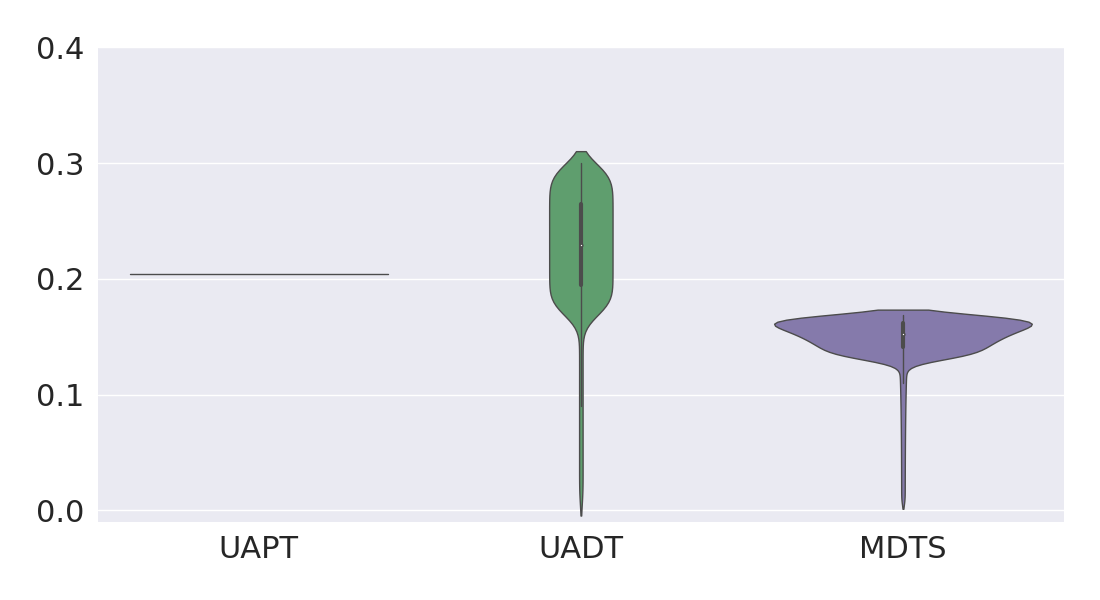}
    \vspace{-2mm}
    \caption{Uncertainty along 100 runs of 2500 seconds each.}
    \label{fig:box-plot_uncertainties}
  \end{figure}

In order to better characterize the reduction of the error in the mitigated approach, we additionally measured the uncertainty of the box temperature for the 3 approaches: UAPT, UADT and MDTS. In other words, we measured $U$, $U^\prime$ and $\mu$ (see Figure~\ref{fig:mitigated_uncertainty_by_dt}). The measures were realized along 100 runs of 2500 seconds and logged at each time step. The results are represented in Figure~\ref{fig:box-plot_uncertainties} using another violin plot. The figure shows the distribution of the values obtained for $U^\prime$ (from UAPT), $U$ (from UADT) and $\mu$, the mitigated uncertainty. 
We can see that, as expected, the uncertainty from the measurements $U^\prime$ remains constant all over the runs. The uncertainty from the model $U$ is spread between a very small one (0.005 at the first simulation step of each run) to the maximum we defined for its reliability, \ie 0.3 in this experiment. The mitigated uncertainty $\mu$ is significantly smaller than both $U$ and $U^\prime$. This explains the reduction of the error observed in Figure~\ref{fig:box-plot_error_to_measurand}. 

\subsection{Managing Inconsistent Behaviors}\label{sec:managingInconsistentBehaviors}

Sections~\ref{sec:consistentBehaviors} and \ref{sec:inconsistentBehaviors} defined the notions of consistent and inconsistent behaviors of the two twins. Inconsistent behaviors may result in divergent executions, which may require reactions to reverse them or react to their consequences.  

There are several ways to understand and react to the divergent behaviors of the two twins, depending on the specific system and the application domain. At best, {inconsistency} can be tolerated if {it is} considered harmless or caused by temporary situations. Whatever the case, inconsistent behaviors should always be detected because of the damaging, or even dangerous, consequences they may entail.  

For example, occasional inconsistency may reveal intermittent small deviations, or simply that the uncertainty of one of the twins is not correctly assessed. 
Detecting when these inconsistencies occur can help engineers understand the source of the problem.

For instance, considering the incubator, if inconsistencies always occur around the top of the box temperature time series after the heater is \textit{off} and before the temperature decreases, this may be due to poorly quantified sources of uncertainty in the heat restitution capacity of the heater, explaining why heater inertia is different between twins.

It may also be important to monitor the frequency of occasional inconsistencies, as the accuracy of some sensors decreases over time. An increase in the frequency of inconsistencies may reveal, \eg the aging of a sensor, possibly requiring an adjustment or predictive maintenance. 

Finally, inconsistencies lasting for several successive values may reveal deep divergences between the twins. Such divergences may be due to a problem or failure in one of the twins. In the PT, it may be due, \eg to the breakage of a physical part or to an unexpected event in its environment. In the DT, divergences may be due to a bug in the solver, a too-abstract model, or even to a source of epistemic uncertainty that has not been considered~\cite{OBERKAMPF2002333}. 

In all cases, the user {(and the services)} of the DTS must be warned since a proper reaction is often required. 
Furthermore, if the controlled process is critical and the source of divergence is not yet understood, a safety procedure could be initiated to bring the controlled process into a safe state. This can be difficult if we consider complex processes, such as autonomous vehicles, or more straightforward if we consider the incubator, where the heater can simply be turned {off}.

\begin{figure}
    \centering
    \includegraphics[width=.9\columnwidth]{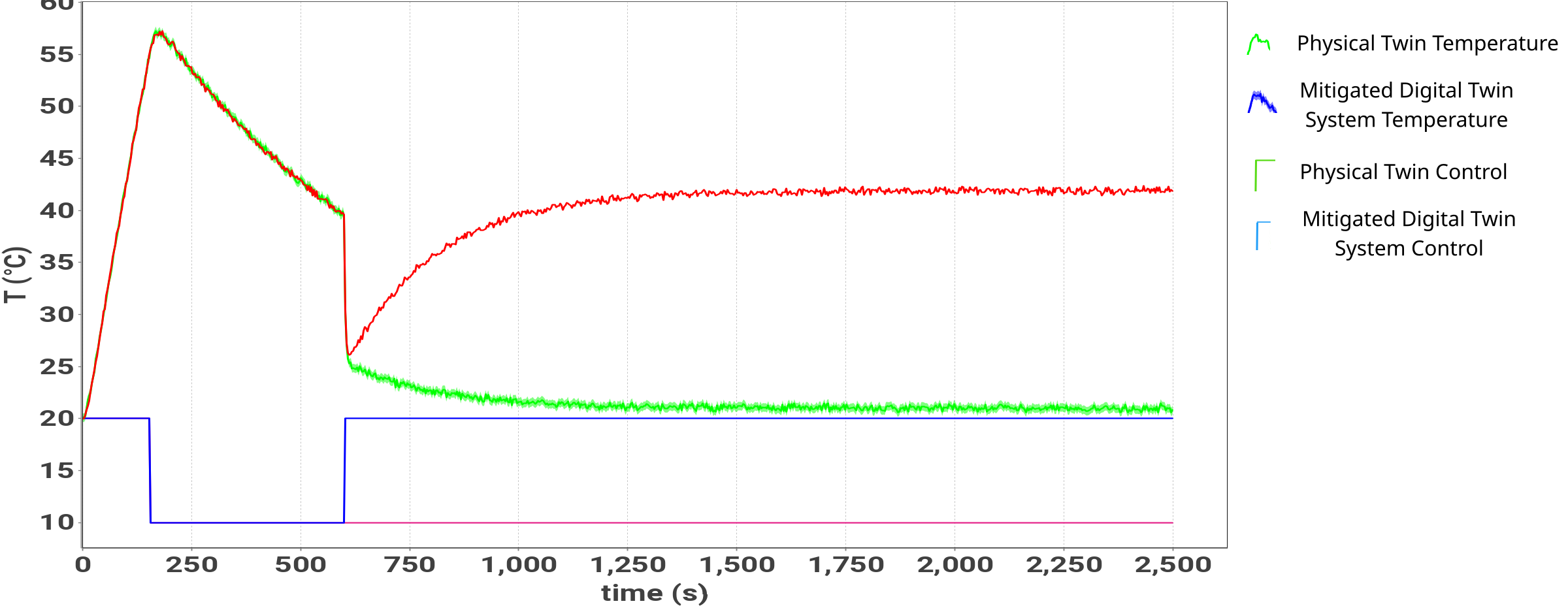}
    \caption{Consistency Based Detection of a Failure introduced at time 600 seconds. By using the proposed notion of consistency, the proposed mitigated digital twin system successfully detects the failure and put the system in safe mode. The traditional physical twin continues trying to achieve the requested control, possibly leading to other failures. }
    \label{fig:failureDetection}
  \end{figure}

To illustrate the use of this consistency checking, we introduced an artificial unexpected event during the simulation {after 600 seconds of incubation}. This event drastically reduces the insulation of the incubator (\eg due to the lid opening or box breakage). 
Our approach promptly detects this anomaly by checking the consistency between the physical and digital twins and stops the system safely by turning off the heater. {In contrast}, the classical approach, which lacks any anomaly-detection mechanism, fails to detect this divergence and {tries} to maintain the desired temperature, resulting in potentially harmful consequences. It could be possible to add an ad-hoc anomaly detection mechanism in the classical approach. However, we believe that using our notion of inconsistency between the PT and the DT behaviors provides a natural support for this task.

Figure~\ref{fig:failureDetection} illustrates the effectiveness of our inconsistency approach for anomaly detection. Few control cycles after the lid is opened, our approach successfully detects the divergence between the twins based on their uncertainty. {The incubator heater is stopped and} the temperature gradually decreases toward room temperature over the next 2000 seconds (approximately 33 minutes), \ie until the end of the simulation. {Conversely}, in the classical approach, the controller keeps the heater on during these 33 minutes in an attempt to maintain the desired temperature, resulting in a significant waste of energy {and potential danger for the incubated object and its environment}. 

\subsection{Limitations of our proposal}

So far, we have discussed the features and advantages of our proposal. This section discusses some of the limitations that we have also identified. 

One of the main limitations is that the exploratory phase of the proposed technique has been conducted only on the case study of the incubator. This does not ensure that the technique will be fully applicable to other contexts or systems with different characteristics. However, based on the results, we believe that the technique could be applicable to any DTS that operates by synchronizing the behavior of both twins.

Furthermore, the tests have been conducted mainly on synthetic data, which means that we do not know of other challenges that may arise when applying the technique in practice. For example, unknown uncertainties that we may have abstracted in the synthetic model, or emergent interactions between them~\cite{dur36061}, could also influence the system behavior. We however tried to use synthetic data that was as realistic as possible. 

Another limitation may lie in the quantification of the initial uncertainty of the parameters. Each simulation parameter has an associated initial level of uncertainty that will then propagate. Determining the value of that uncertainty may not be trivial because it may not be directly derived from the measurement device, or it may be a value that changes over time (\eg sensor sensibility decay). Accurately calibrating these values is essential to obtain satisfactory results.

Despite these limitations, our proposed technique has shown promising results in detecting anomalies and improving the {behavior} of DTS. With further research and testing, we believe that this technique can be adapted and applied to a variety of systems, leading to significant benefits in terms of uncertainty mitigation.

\section{Related work}
\label{sec:Relatedwork}

There are three main lines of work closely related to our proposal. The first one concerns the uncertainty in simulation models of control systems. The second one is related to the explicit representation of uncertainty in software models. The third line deals with uncertainty in the twinned systems. 

\subsection{Uncertainty in control systems}

Uncertainty in control systems and their simulation has been traditionally managed using different approaches. 

One approach uses intervals to represent the possible values of uncertain attributes. For example, Fujimoto~\cite{Fujimoto99,LoperF00} use time intervals to deal with the concepts of approximate time and event ordering in the context of DEVS~\cite{DEVS}.  
Saadawi and Wainer also explored replacing time datatype in DEVS models by intervals in their RTA-DEVS  formalism~\cite{RTA-DEVS}. Other proposals provide methods to specify uncertainty in the state, input, and output variables in addition to the time variable~\cite{VicinoWD22,Dorato1987}. Works such as~\cite{KobayashiSHCIK21} make conservative decisions based on intervals to robustify the specification of controllers of cyber-physical systems so that they satisfy safety requirements under uncertain conditions. 
Reachability is the set of techniques for quantifying and propagating intervals~\cite{Donze2010}. Reachability analysis can be used for interval uncertainty quantification from noisy data and its propagation~\cite{Wright2022}. 

However, specifying and operating with intervals requires a significant effort by the modeler since there is no direct support for making computations with them, such as arithmetic operations or comparisons, which are burdensome and error-prone tasks. In addition, intervals provide a too coarse-grained and pessimistic representation of uncertainty, which results in very conservative (also called \textit{cautious}) simulations~\cite{Wittenmark75}.

Another set of papers studies the relationship between the uncertainty of the input parameters and that of the simulation results, in order to define measures for risk quantification under input uncertainty. In general, there are two sources of uncertainty in a typical stochastic simulation experiment: the extrinsic uncertainty on input parameters and the intrinsic uncertainty on output response (also called stochastic uncertainty) which reflects the inherent stochasticity of the system. 
Some authors~\cite{Gordy2010,Zhu2020} propose nested Monte Carlo simulation approaches
to estimate them~\cite{JCGM101:2008}. Other set of works use the Kalman Filter for uncertainty mitigation and propagation~\cite{Mezic2004}. However, in case of divergence, the Kalman filter will internally decide whether it believes the physical data or the model and continue to average both values. Others~\cite{Cheng2004} propose statistical methods for the calculation of confidence intervals for the mean of a simulation output. They get more accurate results than those that use interval arithmetic or very conservative (\ie robust) estimations.  However, both the complexity of the calculations and their computational costs hinder their applicability. Finally, recent works such as~\cite{JV2023} make use of random variables to represent uncertain attributes and uncertainty propagation is achieved using closed-form solutions mitigate these problems. In our work, we follow this latter approach.

\subsection{Uncertainty in software models}

The explicit representation of uncertainty is a well-known challenge. The survey~\cite{TroyaMBV21} covers current approaches, for which significant challenges still remain. For example, there are very few programming or modeling language libraries that support measurement uncertainty, and even those that support uncertainty propagation (e.g.,~\cite {PythonUncertainties,SoerpPythonUncertainties,OpenTURNS,UPSoftware}) do not provide the performance required to run simulations and do not support comparison between uncertain numbers. Uncertain comparisons are critical for adaptive systems since their decisions must be based on comparisons. In physical systems, logical variables representing comparisons between quantities rarely result in crisp true or false values. Instead, extensions of Boolean logic such as probability theory~\cite{Finetti2017} are more appropriate. In this paper, we have used the Java library described in~\cite{2020-Bertoa-SoSym-j} that supports all basic primitive data types endowed with uncertainty. 

\subsection{Uncertainty in Digital Twins}

The treatment of uncertainty in the domain of digital twins is gaining attention. Recent works aim at identifying the uncertainties of relevance in this domain, either in general~\cite{KulkarniBCB22, Oquendo19, abs-2208-12904} or focusing on particular application domains such as energy~\cite{abs-2303-10954}, industry 4.0~\cite{Xu00A22,LuoTD21} or automotive~\cite{abs-2103-03680}. More specialized works, such as \cite{Woodcock2021}, discuss how uncertainty can be quantified and propagated to monitor safety properties using monte-carlo simulations. 
These works tend to focus on system or model uncertainties, and how the uncertainty of one can be mitigated by that of its twin counterpart. To the best of our knowledge, ours is the first work that takes into account both system and model uncertainties, recognizes that they are different in nature, and uses them in combination to reduce the overall system uncertainty.

\section{Conclusions}
\label{sec:Conclusions}
In this paper, we have shown how the explicit representation of the sources of uncertainty of both the system and the models of a DTS can be used to improve its behavior. Furthermore, this enables a more accurate comparison of the behaviors of the physical and the digital twins. It also provides support to assess their validity and determine when the two behaviors are consistent or, on the contrary, diverge. 

Our work can be continued along several lines. First, further experiments with different types of DTS will allow us to gain more confidence in the applicability and effectiveness of our proposal, identify possible limitations, and improve it. Second, dealing with other types of uncertainties, \eg epistemic ones, represents an interesting research challenge. We also plan to use our approach within a validity framework for analyzing both models and experiments.

\textit{Open research}: All the software, artifacts, and results described in the paper are available from 
 \url{https://github.com/atenearesearchgroup/uncertainty-mitigation-dts}.

\begin{acknowledgement}
  This work was initiated during the $18^{th}$ CAMPaM workshop supported by Università di Corsica, Antwerpen University, and the Université Cote d'Azur. 
\end{acknowledgement}

\begin{funding}
  It was partially funded by the DFG (German Research Foundation) – project number 499241390 (FeCoMASS) – and "Kerninformatik am KIT (KiKIT)" funded by the Helmholtz Association (HGF); the Spanish Government (FEDER/Ministerio de Ciencia e Innovación–Agencia Estatal de Investigación) under projects PID2021-125527NB-I00 and TED2021-130523B-I00; the Poul Due Jensen Foundation, which has supported the establishment of the Center for Digital Twin Technology at Aarhus University; The Natural Sciences and Engineering Research Council (NSERC) under grant RGPIN-2019-05213.
\end{funding}

\printbibliography

\end{document}